\documentclass[prb,showpacs,floatfix,preprint,aps]{revtex4}
\usepackage{epsfig}
\begin{document}
\title
{
The spin-$\frac{1}{2}$ Heisenberg
antiferromagnet on a  1/7-depleted triangular lattice:
Ground-state properties 
}
\author
{
D.Schmalfu{\ss}$^a$, P.Tomczak$^b$,  J.Schulenburg$^a$, J.Richter$^a$
}
\affiliation
{
$^a$Institut f\"ur Theoretische Physik, Otto-von-Guericke Universit\"at Magdeburg \\
P.O.B. 4120, 39016 Magdeburg, Germany \\   
}
\affiliation
{
$^b$Physics Department, Adam Mickiewicz University \\
Umultowska 85, 61-614 Pozna\'n, Poland
}

\begin{abstract}
A linear spin-wave approach, a  variational
method and exact diagonlization are used to investigate the
magnetic long-range order (LRO) of the  spin-$\frac{1}{2}$ Heisenberg
antiferromagnet on a two-dimensional 1/7-depleted triangular  
(maple leaf) lattice consisting
of triangles and hexagons only. This lattice has $z=5$ nearest neighbors and
its coordination number $z$ is therefore between those of the 
triangular ($z=6$) and the kagom\'e ($z=4$) lattices.
Calculating  spin-spin correlations, sublattice magnetization, spin
stiffness, spin-wave velocity   and
spin gap we find that 
the classical 6-sublattice LRO is strongly renormalized by quantum
fluctuations, however, remains stable also in the quantum model.
\end{abstract}


\maketitle


\section{Introduction}

The properties of low-dimensional antiferromagnetic spin systems have been 
subject of many studies in recent years. A lot of activity in this area
was stimulated by the possible connection
of such systems with the phenomenon of high-temperature superconductivity.
But, the rather unusual properties of quantum magnets deserve 
study on their own
to gain a deeper understanding of these quantum many-body systems, 
especially  at low temperatures.
One of the main issues studied is the presence of long-range order (LRO) 
in the ground state
of two-dimensional spin-$\frac{1}{2}$ Heisenberg antiferromagnets (HAF), described
by the Hamiltonian 
\begin{equation}
\label{eq0}
      H = J \sum _{<i,j>}{\bf S}_i \cdot {\bf S}_j
\end{equation}        
on different two-dimensional lattices. 
The sum runs over all pairs of nearest neighbors 
on the lattice under consideration and the coupling $J$ is positive.                            

It is rather well established that LRO is present in the ground state 
of the 
spin-$\frac{1}{2}$ HAF on bipartite lattices  (square \cite{Man}, 
honeycomb\cite{Dea,Mat}, 1/5-depleted square\cite{Tro,Ma}, 
square-hexagonal-dodecagonal\cite{ToRi}) and, contrary to some early 
works\cite{And,Faz},
also on triangular lattice\cite{Ber1,Ber2,Cap}. Those results 
were obtained and confirmed by different methods:
exact diagonalization (ED), Monte-Carlo simulations, spin-wave and variational
approaches, series expansions and others. 
It is also worth noticing that recent experiments show
that real systems can be modeled by spin-$\frac{1}{2}$ Heisenberg antiferromagnets 
with different couplings
on some uniform \cite{Kag,Col} and even depleted lattices \cite{Tan}.

A regular depletion of the triangular lattice by a factor
of 1/4 yields the  kagom\'e lattice with coordination number $z$=4. 
Contrary to the triangular lattice the ground state of the 
spin-$\frac{1}{2}$ HAF 
on the kagom\'e lattice is most likely a spin liquid.\cite{Lech,Wal}
However, the  kagom\'e lattice is not the only regularly depleted 
triangular lattice. 
As recently has been pointed out by Betts \cite{Betts} 
a regular depletion of the triangular lattice
by a factor of 1/7 yields another translationally invariant lattice. 
The coordination number of this lattice is  $z=5$ and
lies between those of the triangular ($z=6$) and the 
kagom\'e ($z=4$) lattices.
According to Betts we will call this lattice in what follows the 
maple leaf lattice.
Since, in general, magnetic order is weakened by frustration
and low coordination 
number $z$, it is natural to 
ask whether the magnetic LRO, present for the HAF on triangular
lattice but absent for the HAF on kagom\'e lattice,
 will survive this 1/7 depletion of the triangular lattice or not. 
In this paper we will study this problem using several  
analytical and numerical methods to calculate the ground state of model
(\ref{eq0}).  

The paper is organized as follows: In Section II we briefly illustrate the
geometrical properties of the lattice and the classical magnetic 
ground state, in Section III exact diagonalization
data for finite lattices of $N=18$ and $36$ spins 
are presented and 
compared with approximate data (spin-wave and variational), 
in Section IV a linear spin-wave approach to this problem is presented,
results of variational calculations are described in Section V, and the
summary is given in Section VI.     

\section{Geometry of the lattice and the classical ground state}

\begin{figure}
\centering\epsfig{file=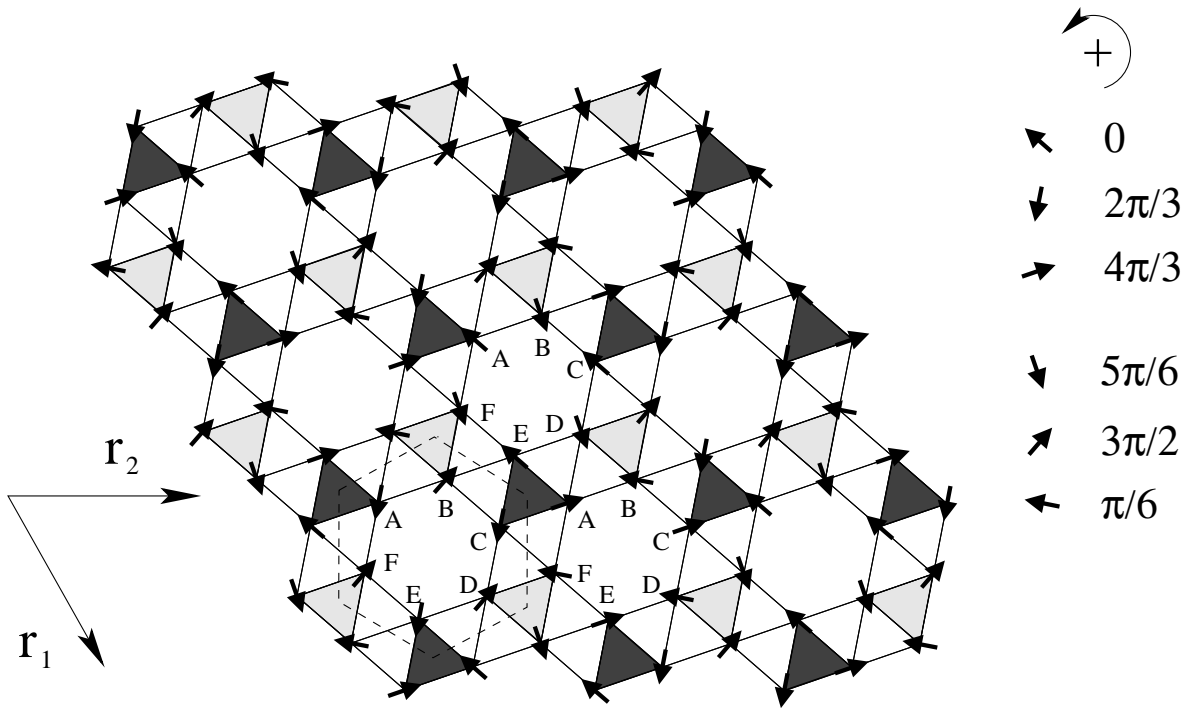,scale=0.6,angle=-90}
\vspace*{-1em}
\caption
{
1/7-depleted triangular (maple leaf) lattice. 
The geometrical unit cell containing six spins is  marked
by the dashed lines. 
This lattice may be split into 6 equivalent triangular sublattices A,B,C, ..., F.
The classical ground state is represented by arrows.
}
\label{fig1}
\end{figure}

The maple leaf lattice is shown in Fig.\ref{fig1}.
It belongs to the class of uniform 
tilings in 2D
built by a periodic array of regular polygons. In each of
the equivalent 
sites 4 triangles and one hexagon meet. 
The maple leaf lattice has no reflection symmetry.
Its unit cell (marked by dashed lines in Fig.\ref{fig1}) 
consists of 6 sites and 15 bonds. 
The underlying Bravais lattice is a triangular one. The basis
vectors are ${\bf r}_{1}=b\sqrt{7}\left[\sqrt{3}/2,1/2 \right]$ 
and  
${\bf r}_{2}=b\sqrt{7}\left[0,1\right]$, where $b$ is the distance 
between neigboring sites. 
More information can be found in Ref. \cite{Betts}.

The ground state of a classical spin system on such 
a lattice 
forms 
the starting point for the calculation of the ground state 
properties of the quantum HAF within the spin-wave method (section IV)  
and variational method (section V). 
As reported previously\cite{Sch}, this ground state is a non-collinear
(canted) planar state with six sublattices. 
It can be characterized as follows:
We denote the position of $i$-th hexagon (unit cell) by  lattice vector 
${\bf R}_{i}$
and  label the sites in the unit cell by the
running index $n=1,...,6$. Then we can write 
\begin{equation} 
{\bf S}_{in}=s\left(\cos\left(\phi_{n}+{\bf Q}\cdot{\bf R}_{i}\right){\bf e}_{1}+
\sin\left(\phi_{n}+{\bf Q}\cdot{\bf R}_{i}\right){\bf e}_{2}\right),
\label{eq1}
\end{equation}  
where ${\bf e}_{1}$ and ${\bf e}_{2}$ are arbitrary orthogonal unit vectors.
For the angles $\phi_n$ we have \\
$\phi_{n}-\phi_{m}=\pm \alpha$ 
for nearest neighbors on the hexagon within the unit cell. 
The corresponding product of two spin vectors reads
$
{\bf S}_{in}\cdot{\bf S}_{mj}=s^{2}\cos\left( \phi_{n}-\phi_{m}+{\bf Q}\cdot\left({\bf
R}_{i}-{\bf R}_{j}\right)\right).
$
The classical ground state corresponds to two sets of 'wave vector' {\bf Q} 
and pitch angle $\alpha = \phi_{B}-\phi_{A}$, namely 
${\bf Q}_{1}=2\pi \frac{1}{b\sqrt{7}}\left[\frac{1}{\sqrt3},\frac{1}{3} \right], 
\alpha_1=-\frac{5}{6}\pi \;$ 
and
$\; {\bf Q}_{2}=2\pi \frac{1}{b\sqrt{7}}\left[ 0,\frac{2}{3} \right], 
\alpha_2=\frac{5}{6}\pi$. 
This is a kind of trivial degeneracy which one can also encounter
in the system of classical spins residing on the triangular lattice.
The classical ground state energy is given by
\begin{equation}
E_{0}^{cl}=-\frac{J}{2}Ns^{2}\left(1+\sqrt{3}\right),
\label{eq4}
\end{equation}
where $N$ is the number of sites.                              
For spin $s=\frac{1}{2}$ and $J=1$, we have the energy per bond 
$E/bond = -(\sqrt3+1)/20 \approx -0.137$. Notice that for the HAF
on triangular and kagom\'e lattices $E/bond = -0.125$.

The classical ground state for 
$\; {\bf Q}={\bf Q}_{2}$ is illustrated in Fig.\ref{fig1}. 
One has six triangular sublattices A,B,...,F.
The classical spins attached to the sublattices 
are  rotated from one unit cell to the next one  by the angle  
$-\frac{2}{3} \pi$ 
in the direction of basis
vector ${\bf r}_{1}$ 
and by the angle  $\frac{2}{3} \pi$ in the direction of 
${\bf r}_{2}$.
The angle between the
nearest spins on each hexagon  is $\frac{5}{6} \pi$  or  $-\frac{5}{6} \pi$ 
and three spins residing on each equilateral triangle 
(marked by light and dark grey in Fig.\ref{fig1}) 
coupled to three nearest hexagons
form a $120^\circ$-structure. 

Though the classical ground state
of the maple leaf lattice is more complex than that of the
triangular lattice both are N\'{e}el states, however, with more than two
sublattices. At the same time the classical ground state properties 
of the kagom\'e
lattice exhibiting a nontrivial ground state degeneracy 
are completely different.

\section{The exact diagonalization}
 
\begin{figure}[ht]
\centerline{
\epsfig{file=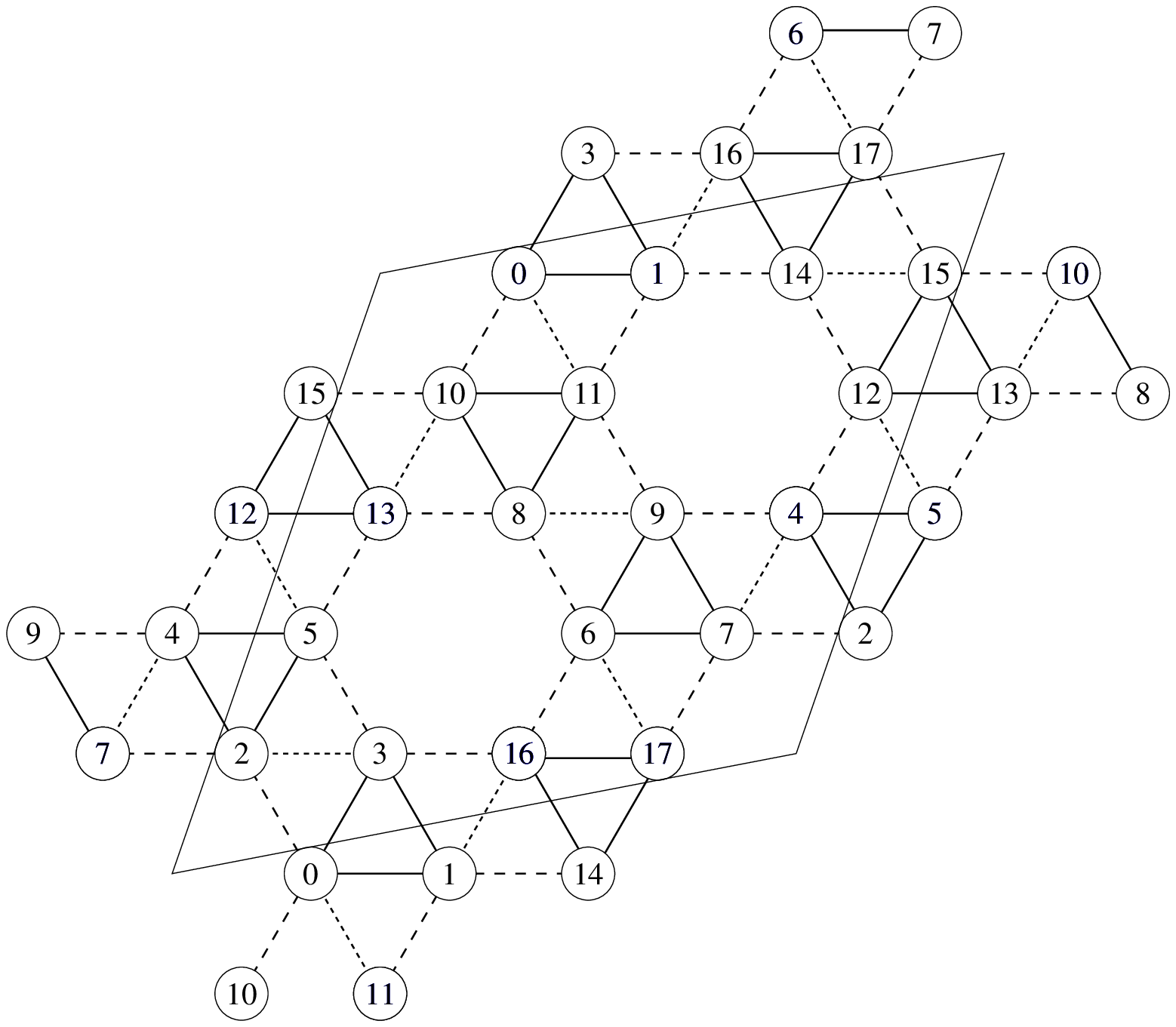,scale=0.66,angle=0}}
\vspace*{-5.1cm}
\centerline{
 \epsfig{file=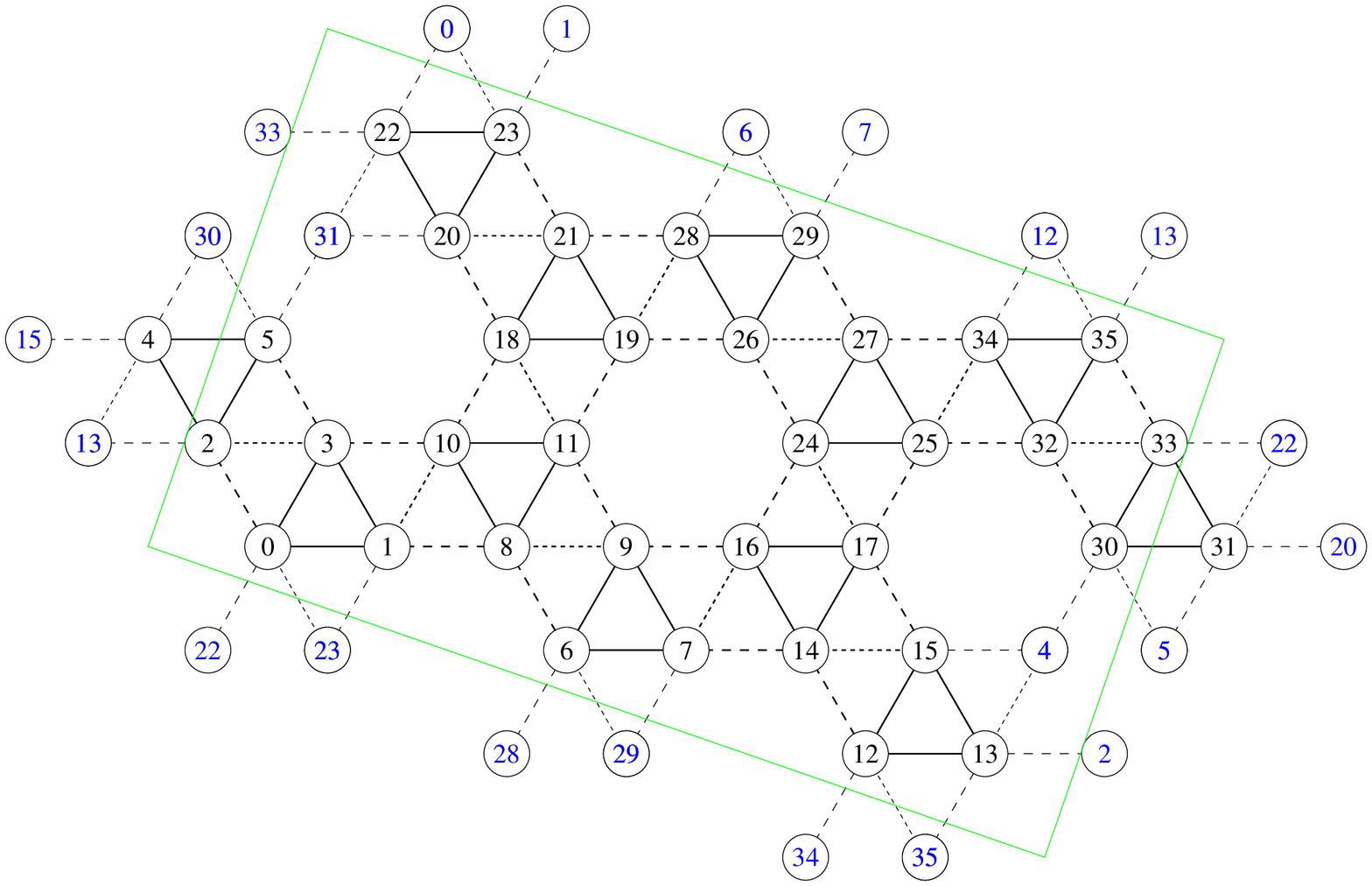,scale=0.66,angle=-90}}
\vspace*{-1.2cm}
\caption{ Finite  
1/7-depleted triangular (maple leaf) lattices with $N=18$ and $N=36$ sites.}
\label{fig2}
\end{figure}

We used the Lanczos algorithm to calculate the lowest
eigenvalue and the corresponding  eigenstate of finite lattices 
of $N=12,18,24,30,36$
sites. This method has sucessfully applied to finite triangular and 
kagom\'e lattices.
Unfortunatly, for the maple leaf lattice only the lattice with $N=18$ has 
the complete  p6-symmetry of 
the infinite lattice and only multiples of 18 fit to the symmetry of the
classical ground state. Hence we focus on the lattices with 
$N=18$ and $N=36$  shown in Fig.\ref{fig2}. 
To reduce the Hilbert space of the Hamiltonian
we use all possible translational and point symmetries
as well as spin reflection.
The number of symmetries of the $N=36$ lattice is lower than
that of corresponding $N=36$ triangular and 
kagom\'e lattices and the number of symmetrized basis 
in the $S_{tot}^{z}=0$ ground state sector is 378,221,361.
The ground state energy per bond for $N=18$ is $E_{0}/bond=-0.2190042 J$
and for $N=36$ is
$E_{0}/bond=-0.2155890 J$.  
The spin-spin correlation functions for $N=18$ are collected in Table
\ref{tab1} 
and for $N=36$ in Table \ref{tab2} where they are compared to those obtained 
within spin-wave and variational approach (see below).
In the fully symmetric $N=18$ lattice we have three different 
nearest-neighbor (NN) correlations.  
The NN correlation $\langle {\bf S}_0  {\bf S}_1 \rangle = -0.186299$ 
(solid lines in Fig.\ref{fig2})
corresponds to the classical $120^\circ$ bond (see Fig.\ref{fig1}); 
the respective averaged value 
for $N=36$ is $-0.177732$. Both values are very close 
to the NN correlation for the triangular lattice. The NN correlation
along a hexagon (dashed lines in Fig.\ref{fig2}) is strong 
$\langle {\bf S}_0  {\bf S}_2 \rangle = -0.366673$ 
(the respective averaged value for
$N=36$ is $-0.365555$) and is close to the NN correlation of the honeycomb 
lattice. Finally, the NN correlation corresponding to a 
classical $90^\circ$ bond (dotted lines in Fig.\ref{fig2})
is very small
$\langle {\bf S}_0  {\bf S}_{11} \rangle = 0.010923$ 
(the respective averaged value for $N=36$ is $-0.037491$).
Hence the NN correlations of the quantum system reflect very well 
the classical ground state.   

The finite-system order parameter corresponding to the classical ground state 
is the structure factor (square of sublattice magnetization)
\begin{equation}
\label{order}
 m^2=\frac{6}{N^2}\sum_{i,j=1}^{N/6} \sum_{n=1}^6 e^{i{\bf Q}({\bf R}_i-{\bf R}_j)}
      \langle{\bf S}_{in}{\bf S}_{jn}\rangle.
\end{equation}
The values  for $N=18$ and $36$ are listed in Table \ref{tab4}.
A finite-size scaling of the order parameter  
with only two points seems to be not reasonable, however, doing so with a
$N^{-1/2}$ scaling we obtain a finite value of $m^2$ for $N \to \infty$.

\begin{figure}[ht]
\centerline{
\epsfig{file=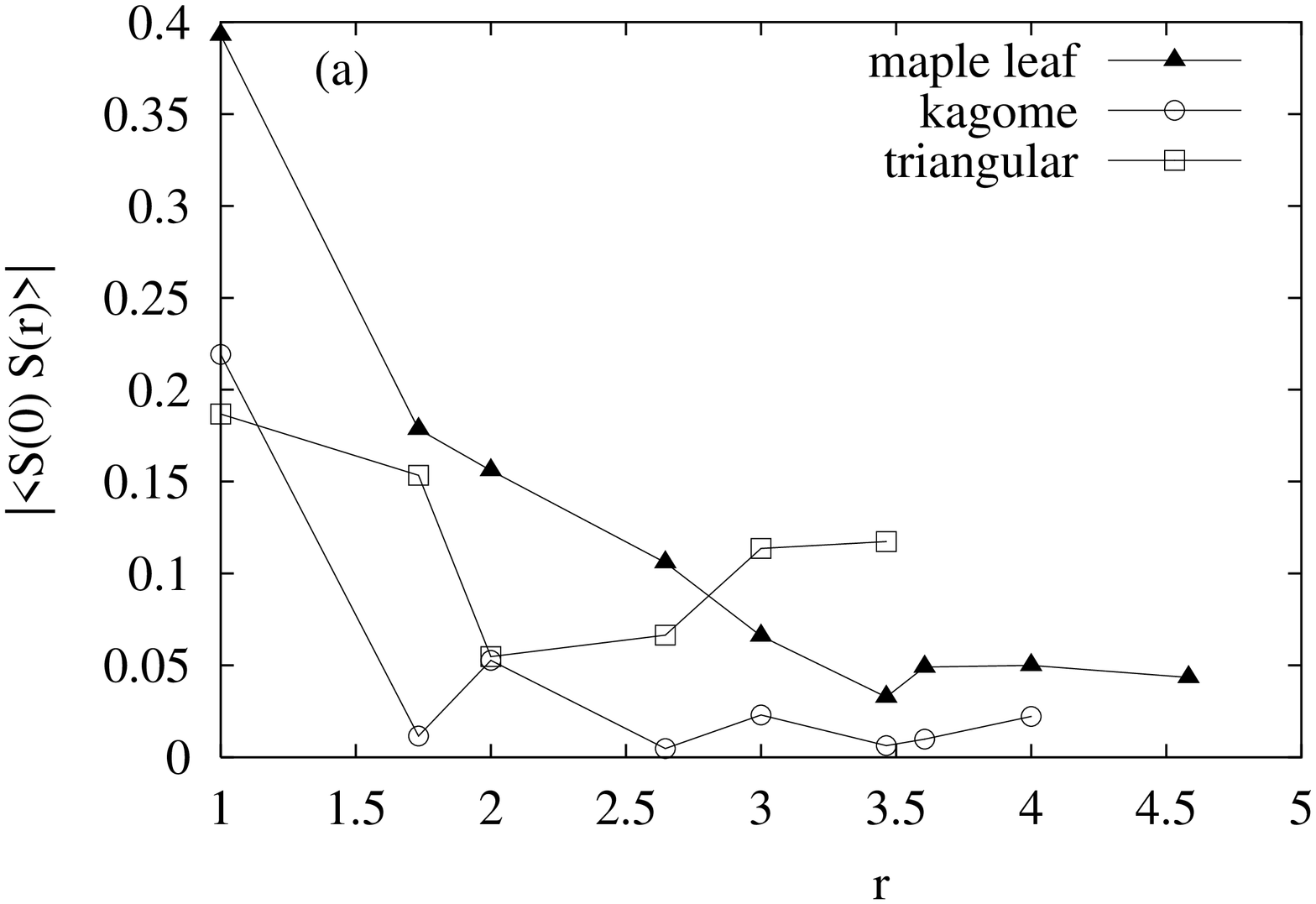,scale=0.58,angle=-90}}
\vspace*{-.1cm}
\centerline{
\epsfig{file=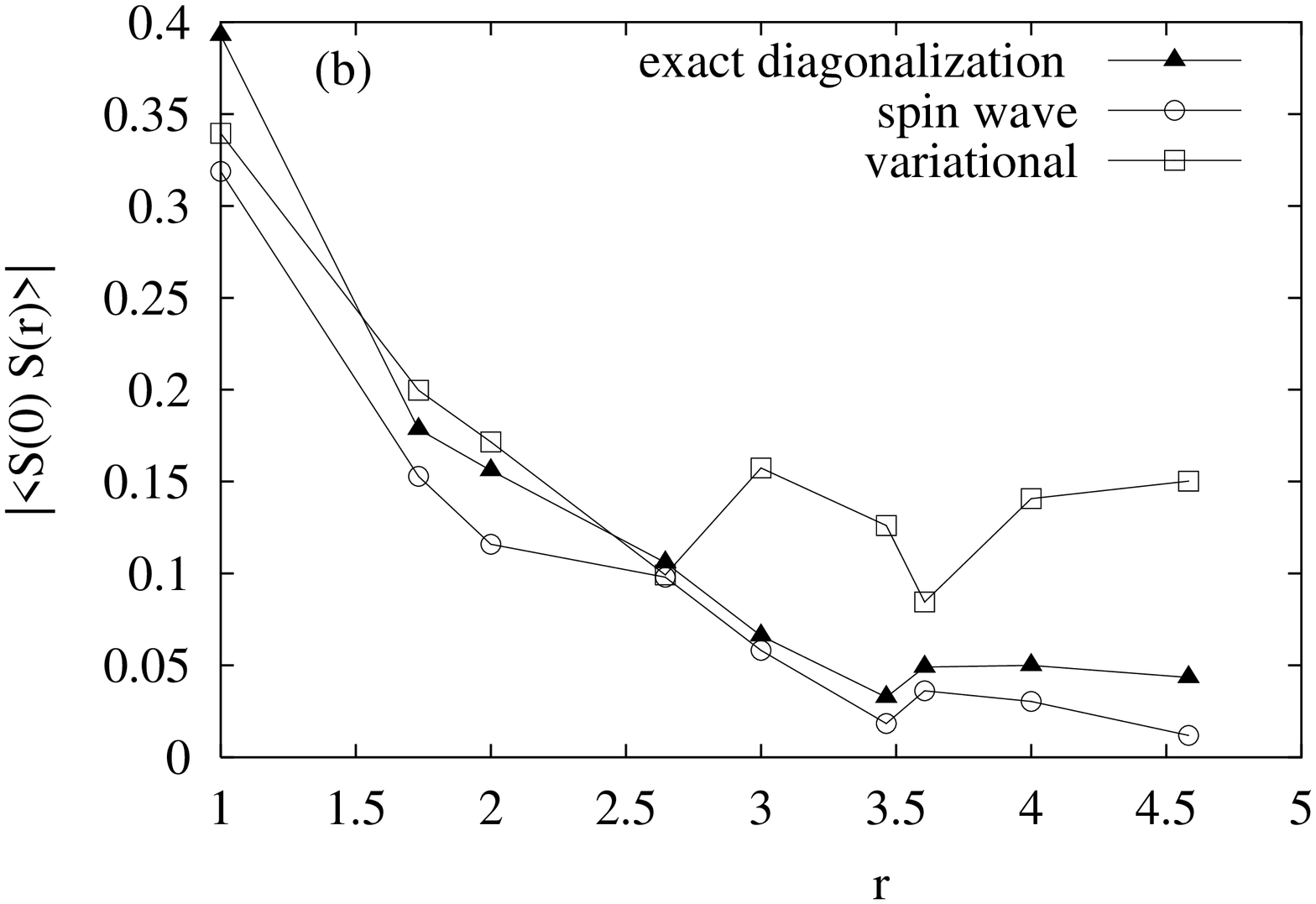,scale=0.58,angle=-90}}
\vspace*{1em}
\caption{ The dependence of the spin-spin correlation 
on the Euclidean
distance for the HAF on the finite maple leaf lattice with $N=36$ sites  
shown in  Fig.\ref{fig2}. (a) comparison
between exact diagonalization results for different lattices.
Data for  kagom\'e are taken from Ref. \cite{Leu}.
(b) Comparison
between exact diagonalization, spin-wave and variational results.
The lines are guide for the eyes. }
\label{fig3}
\end{figure}

A better way is the direct comparison of the spin-spin 
correlations with 
those for the HAF on triangular and kagom\'e lattices\cite{Leu}.
For the presentation of the data we have to take into account that in
the classical six-sublattice N\'{e}el 
state we have for instance spins  with a relative angle of 
$90^\circ$ leading to special correlations being
zero for arbitrary distances. Therefore we consider as a measure for
magnetic order the strongest correlations. Consequently we present 
in Fig.\ref{fig3}a 
the maximal absolute correlations  
$|\langle {\bf S}_0  {\bf S}_{r} \rangle|$  versus Euclidian distance $r$.
As expected we have 
very rapidly decaying correlations for the disordered 
kagom\'e case, whereas the correlations for the N\'{e}el ordered  
triangular lattice are much stronger  for larger distances.
Though the correlations for the maple leaf lattices are smaller than those of the
triangular lattice they are significantly stronger than those of  the 
kagom\'e lattice and a kind of saturation for larger distances is suggested.

The results for the spin-spin correlation 
are used to estimate the quality of
the spin-wave and variational method used below 
by comparing the exact and approximate  correlations  
$|\langle {\bf S}_0  {\bf S}_{r} \rangle|$ 
for the finite lattice of
$N=36$ (Fig.\ref{fig3}b).

\section{The linear spin-wave approach}

Taking into account that we have six sites in the geometrical unit cell the 
appropriate representation of the general Hamiltonian (\ref{eq0}) reads
\begin{equation}
\label{eq0a}
      H = J \sum_{<i,j;n,m>}{\bf S}_{in} 
\cdot {\bf S}_{jm}
\end{equation}        
where $i,j$ label the unit cells and $n,m=1,\ldots, 6$ the different sites
in one unit cell. Of course,  
the sum runs over neighboring sites, only. 
The linear spin-wave theory (LSWT) is carried out as usual. However, we need
at least six different types of magnons, which makes the calculation more
ambitious than for the triangular or the kagom{\'e} lattice.   
We use as quantization axis the
local orientation of the spins in the classical ground state. 
Performing  
the linear Holstein-Primakoff transformation 
the 
scalar product ${\bf S}_{in} \cdot {\bf S}_{jm}$ in (\ref{eq0a})) is replaced 
by the bosonic quadratic form
\begin{eqnarray}
{\bf S}_{in}{\bf S}_{jm}&\rightarrow &
s^{2}\cos\Theta_{nm}^{ij}-s\cos\Theta_{nm}^{ij}\left(a_{ni}^{+}a_{ni}+a_{mj}^{+}a_{mj}\right)\nonumber\\
&+&s\left(\cos\Theta_{nm}^{ij}-1\right)\left(a_{ni}^{+}a_{mj}^{+}+a_{ni}a_{mj}\right)/2\nonumber\\
&+&s\left(\cos\Theta_{nm}^{ij}+1\right)\left(a_{ni}^{+}a_{mj}+a_{ni}a_{mj}^{+}\right)/2,
\label{eq5}
\end{eqnarray}
where $n,m=1,\ldots, 6$ label the
different magnons in a unit cell $i$. $\Theta_{nm}^{ij}$ represents 
the angle between the respective classical spin vectors. 
After transforming the Hamiltonian (\ref{eq0a}) 
into the ${\bf k}$-space one obtains
\begin{equation} 
H=-\frac{J}{2} N s^{2}\left(1+\sqrt{3}\right)+Js\sum_{{\bf k}}H_{\bf k}, 
\label{eq6a}
\end{equation}  
where  
\begin{eqnarray}
H_{{\bf k}}&=&\left(1+\sqrt{3}\right)\left
(a_{1{\bf k}}^{+}a_{1{\bf k}}+a_{2{\bf k}}^{+}a_{2{\bf k}}
+a_{3{\bf k}}^{+}a_{3{\bf k}}+a_{4{\bf k}}^{+}a_{4{\bf k}}
+a_{5{\bf k}}^{+}a_{5{\bf k}}+a_{6{\bf k}}^{+}a_{6{\bf k}}\right)\nonumber\\
&-&\frac {\left(2+\sqrt{3}\right)}{4}\gamma_{1{\bf k}}^{\ast}\left(
a_{2{\bf k}}^{+}a_{3-{\bf k}}^{+}+a_{6{\bf k}}^{+}a_{5-{\bf k}}^{+}\right)
-\frac {\left(2+\sqrt{3}\right)}{4}\gamma_{1{\bf k}}\left(
a_{2{\bf k}}a_{3-{\bf k}}+a_{6{\bf k}}a_{5-{\bf k}}\right)\nonumber\\
&+&\frac {\left(2-\sqrt{3}\right)}{4}\gamma_{1{\bf k}}^{\ast}\left(
a_{2{\bf k}}^{+}a_{3{\bf k}}+a_{6{\bf k}}^{+}a_{5{\bf k}}\right)
+\frac {\left(2-\sqrt{3}\right)}{4}\gamma_{1{\bf k}}\left(
a_{2{\bf k}}a_{3{\bf k}}^{+}+a_{6{\bf k}}a_{5{\bf k}}^{+}\right)\nonumber\\
&-&\frac {\left(2+\sqrt{3}\right)}{4}\gamma_{2{\bf k}}^{\ast}\left(
a_{4{\bf k}}^{+}a_{5-{\bf k}}^{+}+a_{2{\bf k}}^{+}a_{1-{\bf k}}^{+}\right)
-\frac {\left(2+\sqrt{3}\right)}{4}\gamma_{2{\bf k}}\left(
a_{4{\bf k}}a_{5-{\bf k}}+a_{2{\bf k}}a_{1-{\bf k}}\right)\nonumber\\
&+&\frac {\left(2-\sqrt{3}\right)}{4}\gamma_{2{\bf k}}^{\ast}\left(
a_{4{\bf k}}^{+}a_{5{\bf k}}+a_{2{\bf k}}^{+}a_{1{\bf k}}\right)
+\frac {\left(2-\sqrt{3}\right)}{4}\gamma_{2{\bf k}}\left(
a_{4{\bf k}}a_{5{\bf k}}^{+}+a_{2{\bf k}}a_{1{\bf k}}^{+}\right)\nonumber\\
&-&\frac {\left(2+\sqrt{3}\right)}{4}\gamma_{3{\bf k}}^{\ast}\left(
a_{6{\bf k}}^{+}a_{1-{\bf k}}^{+}+a_{4{\bf k}}^{+}a_{3-{\bf k}}^{+}\right)
-\frac {\left(2+\sqrt{3}\right)}{4}\gamma_{3{\bf k}}\left(
a_{6{\bf k}}a_{1-{\bf k}}+a_{4{\bf k}}a_{3-{\bf k}}\right)\nonumber\\
&+&\frac {\left(2-\sqrt{3}\right)}{4}\gamma_{3{\bf k}}^{\ast}\left(
a_{6{\bf k}}^{+}a_{1{\bf k}}+a_{4{\bf k}}^{+}a_{3{\bf k}}\right)
+\frac {\left(2-\sqrt{3}\right)}{4}\gamma_{3{\bf k}}\left(
a_{6{\bf k}}a_{1{\bf k}}^{+}+a_{4{\bf k}}a_{3{\bf k}}^{+}\right)\nonumber\\
&-&\frac {3}{4}\gamma_{1{\bf k}}^{\ast}\left(a_{4{\bf k}}^{+}a_{6-{\bf
k}}^{+}+a_{3{\bf k}}^{+}a_{1-{\bf k}}^{+}\right)-\frac {3}{4}\gamma_{1{\bf k}}\left(a_{4{\bf k}}a_{6-{\bf
k}}+a_{3{\bf k}}a_{1-{\bf k}}\right)\nonumber\\
&+&\frac {1}{4}\gamma_{1{\bf k}}^{\ast}\left(a_{4{\bf k}}^{+}a_{6{\bf
k}}+a_{3{\bf k}}^{+}a_{1{\bf k}}\right)+\frac {1}{4}\gamma_{1{\bf k}}\left(a_{4{\bf k}}a_{6{\bf
k}}^{+}+a_{3{\bf k}}a_{1{\bf k}}^{+}\right)\nonumber\\
&-&\frac {3}{4}\gamma_{2{\bf k}}^{\ast}\left(a_{6{\bf k}}^{+}a_{2-{\bf
k}}^{+}+a_{5{\bf k}}^{+}a_{3-{\bf k}}^{+}\right)-\frac {3}{4}\gamma_{2{\bf
k}}\left(a_{6{\bf k}}a_{2-{\bf
k}}+a_{5{\bf k}}a_{3-{\bf k}}\right)\nonumber\\
&+&\frac {1}{4}\gamma_{2{\bf k}}^{\ast}\left(a_{6{\bf k}}^{+}a_{2{\bf
k}}+a_{5{\bf k}}^{+}a_{3{\bf k}}\right)+\frac {1}{4}\gamma_{2{\bf
k}}\left(a_{6{\bf k}}a_{2{\bf
k}}^{+}+a_{5{\bf k}}a_{3{\bf k}}^{+}\right)\nonumber\\
&-&\frac {3}{4}\gamma_{3{\bf k}}^{\ast}\left(a_{2{\bf k}}^{+}a_{4-{\bf
k}}^{+}+a_{1{\bf k}}^{+}a_{5-{\bf k}}^{+}\right)-\frac {3}{4}\gamma_{3{\bf
k}}\left(a_{2{\bf k}}a_{4-{\bf
k}}+a_{1{\bf k}}a_{5-{\bf k}}\right)\nonumber\\
&+&\frac {1}{4}\gamma_{3{\bf k}}^{\ast}\left(a_{2{\bf k}}^{+}a_{4{\bf
k}}+a_{1{\bf k}}^{+}a_{5{\bf k}}\right)+\frac {1}{4}\gamma_{3{\bf
k}}\left(a_{2{\bf k}}a_{4{\bf
k}}^{+}+a_{1{\bf k}}a_{5{\bf k}}^{+}\right)\nonumber\\
&-&\frac {1}{2}\gamma_{1{\bf k}}^{\ast}\left(a_{5{\bf k}}^{+}a_{2-{\bf
k}}^{+}-a_{5{\bf k}}^{+}a_{2{\bf k}}\right)-\frac {1}{2}\gamma_{1{\bf k}}\left(a_{5{\bf k}}a_{2-{\bf
k}}-a_{5{\bf k}}a_{2{\bf k}}^{+}\right)\nonumber\\
&-&\frac {1}{2}\gamma_{2{\bf k}}^{\ast}\left(a_{1{\bf k}}^{+}a_{4-{\bf
k}}^{+}-a_{1{\bf k}}^{+}a_{4{\bf k}}\right)-\frac {1}{2}\gamma_{2{\bf
k}}\left(a_{1{\bf k}}a_{4-{\bf
k}}-a_{1{\bf k}}a_{4{\bf k}}^{+}\right)\nonumber\\
&-&\frac {1}{2}\gamma_{3{\bf k}}^{\ast}\left(a_{3{\bf k}}^{+}a_{6-{\bf
k}}^{+}-a_{3{\bf k}}^{+}a_{6{\bf k}}\right)-\frac {1}{2}\gamma_{3{\bf
k}}\left(a_{3{\bf k}}a_{6-{\bf
k}}-a_{3{\bf k}}a_{6{\bf k}}^{+}\right)
\label{eq6b}
\end{eqnarray}    
with $\gamma_{{\bf k}n}=\exp(i{\bf k}{\bf q}_{n})$, ${\bf
q}_{1}=b/\sqrt{28}\left(-\sqrt{3},5\right)$, ${\bf
q}_{2}=b/\sqrt{28}\left(-2\sqrt{3},-4\right)$, ${\bf
q}_{3}=b/\sqrt{28}\left(3\sqrt{3},-1\right)$ and with $b$ being the distance
between two neighboring spins. 
This Hamiltonian can be diagonalized 
by the Bogoljubov transformation
\begin{equation}
a_{n{\bf k}}=\sum_{m=1}^{6}u_{n{\bf k}}\alpha_{m{\bf k}}+v_{n-{\bf
k}}^{\ast}\alpha_{m-{\bf k}}^{+}.
\label{eq7}
\end{equation}
The new bosonic operators $\alpha_{m{\bf k}}$ describe the normal modes
$\omega_{m{\bf k}}$. In order to determine them and the
Bogoljubov coefficients one has to solve the following equations
\begin{equation}
\left[\alpha_{m{\bf k}},H\right]_{-}=\omega_{m{\bf k}}\alpha_{m{\bf k}}\quad,\quad 
\left[\alpha_{m-{\bf k}}^{+},H\right]_{-}=-\omega_{m{\bf k}}\alpha_{m-{\bf
k}}^{+}.
\label{eq8} 
\end{equation}
The solution gives six different, non-degenerated spin-wave branches ---
five of them are optical whereas the remaining one is
an acoustical branch. The acoustical branch becomes 
zero in the center (${\bf k}=0$) and at the
edges of the Brillouin zone  (${\bf
k}=\pm{\bf Q}$ with ${\bf Q}=
2\pi \frac{1}{b\sqrt{7}}\left[\frac{1}{\sqrt3},\frac{1}{3} \right]$).
The expansion of the zero modes in the vicinity of those points gives the
spin-wave velocities   
\begin{eqnarray}
c_{{\bf k}=0}&=&Jsb\frac
{\sqrt{14}\sqrt{39+23\sqrt{3}}}{4\left(2+\sqrt{3}\right)},\nonumber\\
c_{{\bf k}=\pm{\bf Q}}&=&Jsb\frac
{\sqrt{7}\sqrt{407+235\sqrt{3}}}{4\left(7+4\sqrt{3}\right)}.
\label{eq9}
\end{eqnarray}        
The acoustical branch of the maple leaf lattice is similar to that 
of the HAF on the 
triangular lattice\cite{Miy,Chu}, where one has a threefold degenerated 
acoustical branch being zero for ${\bf k}=0$ and at the edges of the
Brilloiun zone  $k=\pm{\bf Q}=
\pm 2\pi/a\left[1/\sqrt{3},1/3\right]$.  The
situation for the HAF on the   
kagom{\'e} lattice is completely different. Starting from the so-called
classical ${\bf k}=0$ state one obtains three branches, 
one dispersionsless (flat) mode $\omega_{\bf k}=0$ and
two degenerated acoustical branches \cite{Har}.

In analogy to the triangular lattice\cite{Chu} it can be shown that the zero modes
${\bf k}=0, \pm{\bf Q}$ of the maple leaf lattice describe out-of-plane 
and in-plane oscillations, respectively.
Therefore, we denote
$c_{{\bf k}=0}$ as $c_{\parallel}$ and $c_{{\bf k}=\pm{\bf Q}}$ as $c_{\perp}$.    
Together with the spin-wave velocity the spin stiffness constitutes the
fundamental parameters which determine the low-energy dynamics of magnetic
systems \cite{chakra89}. 
To calculate the spin stiffness $\rho$ in the leading order $s^{2}$
one can use the 
hydrodynamic relation $\rho=\chi c^{2}$. The 
magnetic  susceptibilities $\chi_{\parallel}=\chi_{zz}/V$
(out-of-plane) and $\chi_{\perp}=\chi_{xx}/V=\chi_{yy}/V$ (in-plane) 
can be determined  
minimizing the classical energy in the limit
of a vanishing external field. We find  
$\chi_{zz}=N/J\left(6+\sqrt{3}\right)$,
$\chi_{xx}=\chi_{yy}=2N/3J\left(4+\sqrt{3}\right)$ and $V=7\sqrt{3}Nb^{2}/12$ as 
the volume  of the lattice 
and from the hydrodynamic  relation one obtains
\begin{equation}
\rho_{\parallel}=0.633975Js^{2},\quad \rho_{\perp}=0.211325Js^{2}, \quad 
\rho_{\parallel}/\rho_{\perp}=3 \; .
\label{eq10}
\end{equation}
The comparison with the corresponding parameters calculated in the same
order in $s$ for the  square and the triangular lattice are given in
Table \ref{tab3}.  
We find that the spin stiffness parameters for the maple leaf lattice are
lower than the corresponding values of the triangular lattice indicating
that the N\'{e}el is stronger influenced by quantum fluctuations in the 
maple leaf lattice than in the triangular one.

The ground state energy $E_{0,N}^{sw}$ is given by 
\begin{equation}
E_{0,N}^{sw}=-\frac{J}{2}Ns\left(s+1\right)\left(1+\sqrt{3}\right)
+\sum_{{\bf k}}\sum_{m=1}^{6}\omega_{m{\bf k}}/2,
\label{eq11}
\end{equation}
which leads in the thermodynamic limit to an energy per bond
\begin{equation}
e_{0}^{sw}=\left( -0.5464106s^{2}-0.13652065s \right) J.
\label{eq12}
\end{equation}
The sublattice magnetization
\begin{equation}
\left\langle
S_{in}^{z}\right\rangle_{N}= \left\langle S^{z}\right\rangle_{N}
= s - \frac{6}{N}\sum_{{\bf k}}
\left\langle a_{n{\bf k}}^{+}a_{n{\bf k}}\right\rangle
\label{eq13}
\end{equation}
calculated in the thermodynamic limit is 
\begin{equation}
\left\langle S^{z}\right\rangle_{\infty}=s-0.346.
\label{eq14}
\end{equation}
A comparison between all those values for HAF on square, triangular and 
maple leaf lattices is given in Table \ref{tab3}. 
Obviously, for all these parameters $c$, $\rho$ and 
$\left\langle S^{z}\right\rangle_{\infty}$ the same tendency is found, namely 
to be largest
for the unfrustrated lattice and to be lowest for the frustrated 
maple leaf lattice with $z=5$.   
Notice, that for the  kagom\'e lattice the 
LSWT yields divergent contributions in the sum over $\bf k$ 
in $\left(\langle
S^{z}\rangle_{N} -s\right) \propto  \sum_{{\bf k}}
\left\langle a_{n{\bf k}}^{+}a_{n{\bf k}}\right\rangle$ 
indicating a vanishing sublattice magnetization \cite{asakawa}.

Finally, we compare the spin-spin correlations $\langle {\bf S}_0
{\bf S}_j \rangle $ (where $j$ runs over all spins in the system) 
obtained within the LSWT for
the finite lattice with $N=36$ shown in Fig.\ref{fig2} 
with the exact numerical
Lanczos data (see Fig.\ref{fig3}a and Table \ref{tab2}). 
One finds a surprisingly good
agreement between the approximative  LSWT data and the exact Lanczos data.
Hence the finding of finite sublattice magnetization obtained
within LSWT is supported by the Lanczos data.

\section{The variational approach}
\label{vari}
The classical ground  state, described in Section II,
is the basis for the construction  of the variational
Huse-Elser \cite{HuEl} ground state which,
expanded in the Ising basis states $\mid\!\alpha\rangle$
of the total spin component $S^z=0$  \cite{Car,ToRi},  reads
\begin{equation}
\label{eq17}
\mid\!\!\Psi\rangle =\sum_{\alpha}\,\exp(\frac{1}{2}\tilde{H}_{class}
+\frac{1}{2}\tilde{H}_{quant}
+\frac{1}{2}\tilde{H}_{frust})\mid\!\alpha\rangle . 
\end{equation}
The operators $\tilde{H}$ are  diagonal in the base
$|\alpha\rangle$.
The term
\begin{equation}
\label{eq18}
\tilde{H}_{class} =  -i \sum_j \phi_j S^{z}_j
\end{equation}
produces a proper 'classical' phase for a given
state $|\alpha\rangle$ in the expansion given by  Eq. (\ref{eq17}).
The sum runs over all spins in the system (i.e. the index $j$ corresponds to
a pair $(i,n)$ of indices in Eqs. (\ref{order}) and (\ref{eq0a}).
$\phi_j$ is
the angle specified in Fig.\ref{fig1}.

The second operator containing variational parameters $K_{jk}$
\begin{equation}
\label{eq19}
\tilde{H}_{quant} = \sum _{j,k}K_{jk} S^{z}_jS^{z}_k
\end{equation}
provides
the amplitude for a given basis state $|\alpha\rangle$ and introduces
the quantum corrections
to the classical function  by taking into account spin-spin correlations.
It means that in this approach
one starts from the state with a broken rotational symmetry and this is still
present during the minimization procedure producing the final
symmetry-broken ordered state \cite{HuEl,ToRi}.

Finally, the third operator which contains the variational parameter 
$L_{jkl}$ and the corresponding sign factors $\gamma_{jkl}=\pm 1$
\begin{equation}
\label{eq20}
\tilde{H}_{frust} = i \sum _{j,k,l}\gamma_{jkl} L_{jkl} S^{z}_jS^{z}_kS^{z}_l
\end{equation}
describes an
additional possible change of the classical phase due to the quantum fluctuations.
Following the ideas of  Huse and Elser \cite{HuEl} we assume that the
wave function of the quantum ground state (i.e. $\mid\!\!\Psi\rangle$ from
Eq. (\ref{eq17}) with all three terms $\tilde{H}_{class}$, $\tilde{H}_{quant}$
and $\tilde{H}_{frust}$) 
has the same symmetry properties as its classical part 
(i.e. $\mid\!\!\Psi\rangle$ from
Eq. (\ref{eq17}) with only $\tilde{H}_{class}$):
the sign of the imaginary part of the wave function changes under the rotation
$R_z(\pi/3)$ by the angle $\pi/3$ whereas remains unchanged  under the rotation $R_z(2\pi/3)$
by the angle $2\pi/3$ about the center of
 a hexagon.
This transformation determines the 'shape' of three-spin terms $L_{jkl}$ and 
the proper sign of $\gamma_{jkl}=\pm 1$ in Eq. (\ref{eq20}).
Similarly to the triangular
lattice \cite{HuEl}
the
most simple three-spin terms
are 'dog legs' with $j$ and $l$ being nearest neighbors of $k$. 
For example, for spin number 5 of the $N=18$ lattice shown in  Fig. 2 (top) there 
exist four such 'interactions': 
$L_{3,5,13}, L_{3,5,4}, L_{2,5,12}$ and $L_{4,5,13}$.
Each $L_{ijk}$ is connected with its corresponding $\gamma$ factor, i.e. 
$L_{3,5,13} \to \gamma_{3,5,13}=\gamma_{EFA}$,
$L_{3,5,4} \to \gamma_{3,5,4}=\gamma_{EFD}$, $L_{2,5,12} \to
\gamma_{2,5,12}=\gamma_{BFC}$ and  
$L_{4,5,13} \to \gamma_{4,5,13}=\gamma_{DFA}$, where the letters A,B,C,D,E,F
correspond to the 6 equivalent triangular sublattices illustrated in
Fig.\ref{fig1}.   
Taking into account that $R_z(\pi/3)(ACE)=(BDF)$, $R_z(\pi/3)(BDF)=(CEA)$ 
and $R_z(2\pi/3)(ACE)=(CEA)$, $R_z(2\pi/3)(BDF)=(DFB)$, i.e., the '$120^\circ$ structure'
ACE (dark triangles in Fig. 1) transforms into '$120^\circ$ structure' BDF 
(grey triangles in Fig. 1) under $R_z(\pi/3)$ or 
into itself under $R_z(2\pi/3),$ one obtains a proper signs of 
$\gamma$ factors
\begin{equation}
\label{eq21}
\gamma_{\alpha\alpha\beta} = - \gamma_{\beta\beta\alpha}.
\end{equation}
The index $\alpha\alpha\beta$ means that two (different) spins
in the three-spin term belong to the same '$120^\circ$ structure', 
the remaining one belongs to the other '$120^\circ$ structure'. 
Thus, for example, if one puts $\gamma_{BFC}=1$ it follows that
$\gamma_{CAD} = -1$ (or  $\gamma_{2,5,12}=1$ and $\gamma_{12,13,10} = -1$, 
see Fig. 2). 

How does one choose the variational parameters $K_{jk}$ and $L_{jkl}$
in Eq. (\ref{eq17}) for the HAF
on the maple leaf lattice? We have applied two criteria: a better
choice of parameter space should give a lower value of the ground
state energy and, if two energies for different parameter spaces
are approximately the same, one should  choose the parameter space which
leads a lower 
value of the variance
$\langle H^2 \rangle  -  \langle H \rangle^2 $.
In order to find an
optimal choice of the wave function we have tested some
possibilities for the
parameter space for
 the fully symmetrical $N=18$ lattice  taking into account 
the whole $S^z=0$ basis in
the
expansion (\ref{eq17}).
The best choice found 
 is the following five-parameter space (results for the correlation 
are collected in Table I):
$(K_{hex}, K_{tr}, K_{others}, L_1, L_2)$. Spins 'interacting'
{\em via}
$K_{hex}$ are nearest neighbors lying on
 a hexagon: {\em hex} = AB, BC, CD, ..., FA,
those 'interacting'
{\em via}
$K_{tr}$ are nearest neighbors belonging to
the $120^\circ$ structure: {\em tr} = EC, CA, AE or BF, FD, DB;
and all remaining nearest neighbors are coupled by $K_{others}$,
thus {\em others} = BE, BC, EF.
Note that there is no long-range variational
parameter
for pairs of spins not being the nearest neighbors.
Moreover, 
one takes into account only three, from
the  four existing 'dog leg'
interactions, i.e., 'dog legs' around a hexagon are
absent. 
For example, in each point F one has $L_1 = L_{BFC}$
and $L_2 = L_{EFD}= L_{AFD}$ and $ L_{EFA}$ is absent 
(or correpondingly for spin number 5 in Fig. 2, $L_1 = L_{2,5,12}$
and $L_2 = L_{3,5,4}= L_{13,5,4}$ and  $L_{3,5,13}$ is absent).
All the expectation values of operators reported
in the following
are calculated for this choice of the variational parameters.

\begin{figure}
\centering\epsfig{file=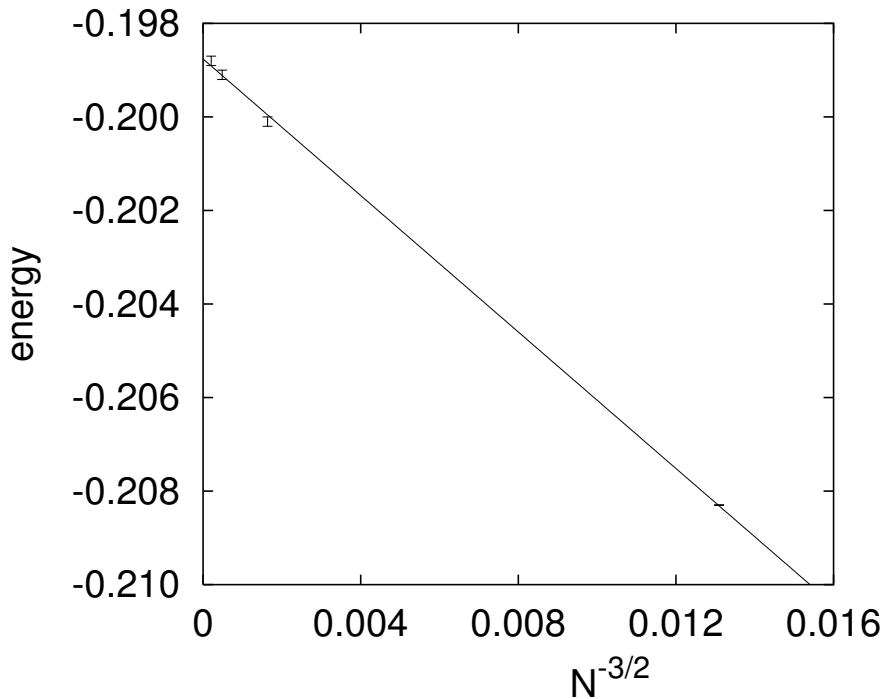,scale=0.6,angle=-90}
\vspace*{1em}
\caption
{
Variational energy per bond for the spin system on the 1/7-depleted
triangular (maple leaf) lattice as a function of $N^{-3/2}$.
}
\label{fig4}
\end{figure}

\begin{figure}
\centering\epsfig{file=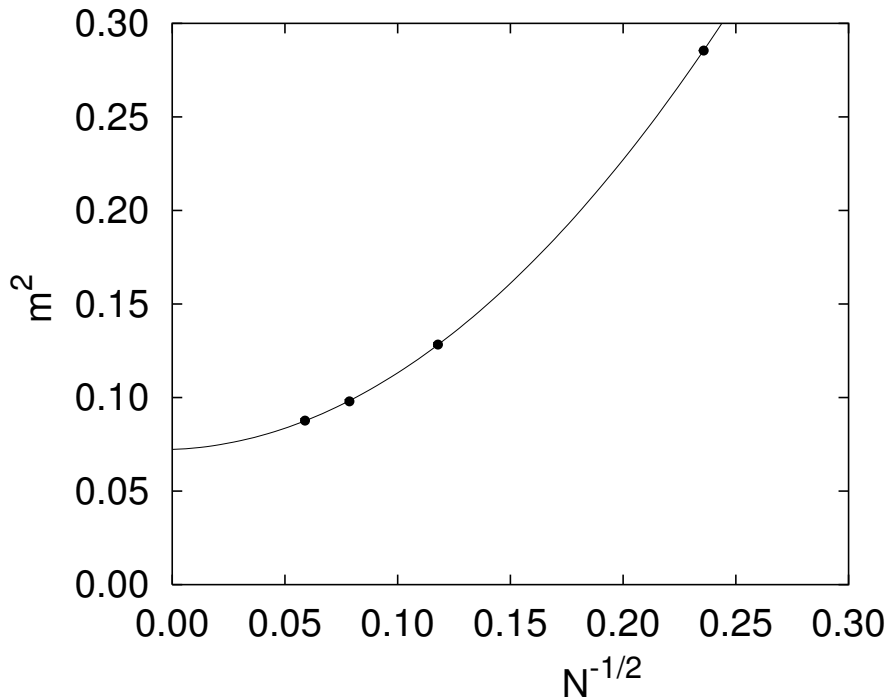,scale=0.6,angle=-90}
\vspace*{1em}
\caption
{
Square of sublattice magnetization, $m^2$, as a function of $N^{-1/2}$.
Circles - values obtained by applying the variational
 method. Solid line -  fit to the circles.
Sizes of circles are comparable to the statistical error bars.}
\label{fig5}
\end{figure}

Having obtained the ground state function
one can calculate the expectation values of the operators
which characterize the ground state of a given, finite spin system.
This can be accomplished by a Monte-Carlo approach\cite{HuEl}
and  the finite size scaling \cite{HaNi} tells how to extrapolate
those expectation values to the thermodynamic limit.
We have investigated  the finite systems of 18, 72, 162, 288 spins
with periodic boundary conditions.
Note that they have the full symmetry of the maple leaf lattice.
The relevant quantities are collected in Table \ref{tab4} and the 
finite size analysis is presented in Figs.\ref{fig4} and \ref{fig5}.

The leading term of the finite-size correction
of the ground state
energy per bond $e$
is $N^{-3/2}$. The data
in 
Table \ref{tab4} can be fitted to this dependence (see Fig.\ref{fig4}) 
and hence the energy
per bond $e_\infty$ in the thermodynamic limit is obtained:
$e(N)=e_\infty+aN^{-3/2}$ with $e_\infty=-0.1988(2) J$ and $a=-0.7327(164) J$.
This value for $e_\infty$ is about 3\% higher than the value obtained from
spin-wave theory (see Eq. (\ref{eq12})).

In Fig.\ref{fig5} the finite-size extrapolation  of the square of 
sublattice magnetization defined in Eq.
(\ref{order}) is shown. We find 
$m^2(N)=m^2_\infty+cN^{-1/2}+dN^{-1}$ with
$m^2_\infty=0.0723(10)$, $c=0.0444(18)$ and $d=3.650(60)$ suggesting
that the long-range magnetic order persists
in the ground state of this spin system. 
Note, however, that the applied variational ansatz tends to overestimate the
magnetic order (see Ref. \cite{HuEl}  and table \ref{tab2} as well as 
Fig.\ref{fig3}a).

\begin{figure}
\centering\epsfig{file=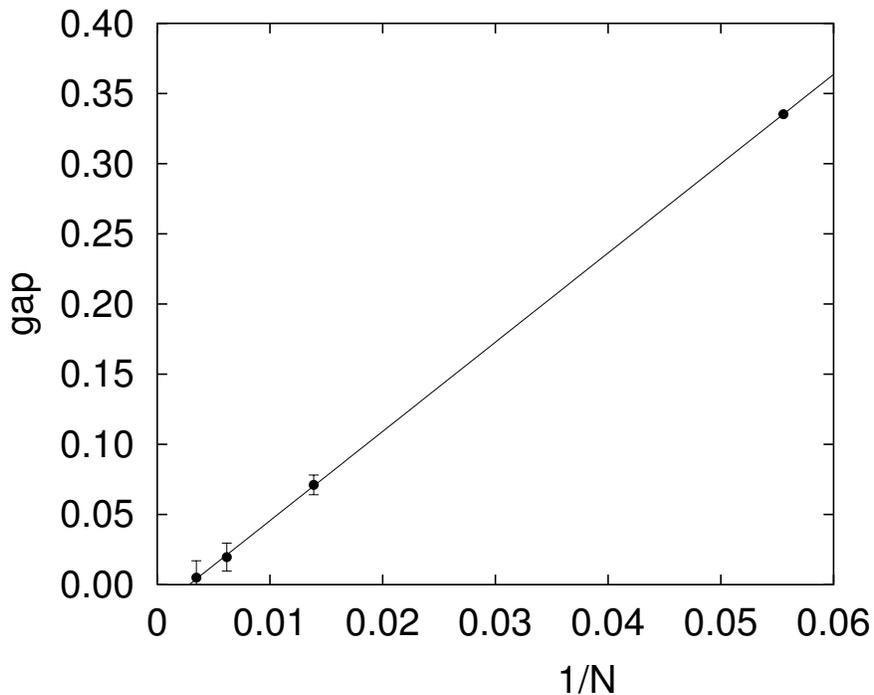,scale=0.6,angle=-90}
\vspace*{1em}
\caption{The spin-gap $\Delta=E_1-E_0$ vs. $1/N$ in units of $J$. 
$E_0$ and $E_1$
denote variational energies of the ground state and a first
excited state, respectively.
Errors result from adding the errors for $E_0$ and $E_1$.
}
\label{fig6}
\end{figure}

The variational approach enables us to calculate the spin
gap $\Delta = E_0 -E_1$, where $E_0$ ($E_1$) is the variational energy in
the subspace of total $S_z=0$ ($S_z=1$). 
This new aspect of the Huse-Elser ansatz 
was  used for the first time in Ref. \cite{ToRi} to calculate the spin gap 
for the HAF
on the square-hexagonal-dodecagonal lattice. Magnetic LRO is connected
with gapless Goldstone modes whereas quantum disorder in the ground state 
is accompanied by a finite  spin gap. Therefore  
the calculation of $\Delta$ yields an additional argument for or against 
the existence
of magnetic LRO order in the ground state.  
Fig.\ref{fig6}  shows
the finite-size extrapolation of the spin gap
according
to
the relation $\Delta(N)=\Delta_\infty+dN^{-1}$
with $\Delta_\infty=-0.0180(10) J$ and $d=6.3610(34)J$.
The negative $\Delta_\infty$ is a result of the limitted accuracy of the
approximation but nevertheless suggests a zero spin gap. Hence we have    
an additional indication for the existence
of LRO.

\section{Summary}
In this paper the results of exact diagonalization, linear spin-wave
theory and a Huse-Elser like variational investigation
for the 
ground state of the spin-$\frac{1}{2}$ Heisenberg antiferromagnet on 
a new 1/7-depleted triangular (maple leaf) lattice are presented.
The coordination number of this frustrated lattice $z=5$ lies 
between those of the triangular and the kagom\'e lattices.
Quantum fluctuations and frustration tend to destroy classical magnetic 
ordering. Their influence becomes the stronger the smaller the coordination
number. But contrary to the kagom\'e  lattice with $z=4$ 
for the maple leaf lattice we find strong arguments that
 the  classical six-sublattice N\'{e}el LRO
survives the strong quantum fluctuations present in this frustrated 
quantum magnet.
This conclusion is drawn from the calculated values of the spin-spin
correlation, sublattice magnetization, spin stiffness, spin-wave
velocity  as well as the spin gap.

The comparison between exact data and approximate data for the spin-spin
correlation on finite lattices 
gives a surprising well agreement
between the linear spin-wave and the exact-digonalization data 
whereas the  variational approach tends to overestimate the 
strength of correlations.   

Finally, we mention that on the  passage from the triangular to the 
1/7 depleted (maple leaf) lattice (i.e., some interactions $J$ in spin
system on triangular lattice are varied from $J=1$ to $J=0$),
one would encounter a transition between three-sublattice
and six-sublattice N\'{e}el LRO which may have interesting features 
worth to be considered in future.\\

{\bf Acknowledgement}

We acknowledge support from the Deutsche Forschungsgemeinschaft 
(Projects No.  Ri 615/10-1 and 436POL 17/5/01)
and from the Polish Committee for Scientific Research 
(Project No. 2 PO3B 046 14).
Some of the calculations were performed at the 
Pozna\'n Supercomputer and Networking Center.

\begin{table}
\caption{ \label{tab1}The exact values of the spin-spin correlation 
$\langle {\bf S}_0 {\bf S}_j \rangle$ compared to their
variational values  for the lattice of $N=18$
sites (see Fig.\ref{fig2}).
}
\begin{ruledtabular}
\begin{tabular}{cccccc}
 $j$ & exact & variational &  $j$ &  exact & variational    \\ \hline
  0 &  0.750000  &   0.750000  &  7 &  0.180027 &  0.200775  \\
  1 & -0.186299  &  -0.180068  &  8 &  0.140873 &  0.162808     \\
  2 & -0.366673  &  -0.343444  &  9 & -0.072868 & -0.106613     \\
  4 &  0.039003  &   0.021877  & 11 &  0.010923 &  0.005425    \\
  5 &  0.145098  &   0.171218  & 17 & -0.174804 & -0.183760   \\
  6 & -0.099672  &  -0.106613  &    &            
\end{tabular}
\end{ruledtabular}
\end{table}

\begin{table}
\caption{The exact values of the spin-spin correlation 
$\langle {\bf S}_0 {\bf S}_j \rangle$   compared to their 
spin-wave and variational values for the lattice of $N=36$
sites (see Fig.\ref{fig2}).
Statistical errors are given in parentheses.
}
\label{tab2}
\begin{ruledtabular}
\begin{tabular}{cccccccc}
$j$ &  exact   & spin-wave  & variational  & $j$  &   exact  & spin-wave  & variational \\ \hline
 1  & -0.1154  &  -0.16759  & -0.1681(90)  &  19  &  0.0660  &   0.05816  &  0.1574(70) \\
 2  & -0.3418  &  -0.31894  & -0.3408(90)  &  20  &  0.1458  &   0.14131  &  0.1603(60) \\
 3  & -0.2008  &  -0.19143  & -0.1703(90)  &  21  & -0.0111  &  -0.00801  & -0.0784(90) \\
 4  &  0.0618  &   0.03135  &  0.0249(70)  &  22  & -0.3929  &  -0.31878  & -0.3395(90) \\
 5  &  0.1394  &   0.15200  &  0.1701(50)  &  23  & -0.0433  &  -0.06302  &  0.0027(60) \\
 6  & -0.0155  &  -0.00625  & -0.0811(90)  &  24  &  0.0434  &   0.01186  &  0.1503(70) \\
 7  & -0.0243  &  -0.01525  & -0.0788(90)  &  25  & -0.0491  &  -0.03622  & -0.0845(90) \\
 8  &  0.0089  &  -0.00304  &  0.0190(60)  &  26  & -0.0448  &  -0.02126  & -0.1393(90) \\

10  &  0.1142  &   0.12745  &  0.1546(60)  &  28  &  0.0034  &  -0.01380  &  0.0089(60) \\
11  & -0.0493  &  -0.04012  & -0.1470(90)  &  29  &  0.0298  &   0.01152  &  0.1249(70) \\
12  & -0.0155  &  -0.00625  & -0.0832(70)  &  30  & -0.1059  &  -0.09804  & -0.0990(90) \\
13  &  0.1488  &   0.12892  &  0.1960(50)  &  31  & -0.0740  &  -0.06787  & -0.0939(80) \\
14  &  0.0327  &   0.01839  &  0.1261(80)  &  32  &  0.0387  &   0.02060  &  0.0156(70) \\
15  & -0.0797  &  -0.06993  & -0.0946(80)  &  33  &  0.1785  &   0.15290  &  0.1997(50) \\
16  & -0.0500  &  -0.03038  & -0.1407(90)  &  34  &  0.0501  &   0.04121  &  0.1313(70) \\
17  &  0.0390  &   0.02791  &  0.0151(60)  &  35  & -0.1561  &  -0.11596  & -0.1716(90) \\
18  & -0.1059  &  -0.09804  & -0.0992(90)
\end{tabular}
\end{ruledtabular}
\end{table}                       
\begin{table}
\caption{Comparison of the LSWT results for the spin-wave velocities, 
spin stiffness parameters and sublattice 
magnetization for the  square \cite{anderson,weihong}, the
triangular\cite{Chu,Miy} and the maple leaf lattice ($J=1$, $b=1$ and $s=1/2$).}
\label{tab3}
\begin{ruledtabular}
\begin{tabular}{cccccc}
lattice     & $c_{\parallel}$  & $c_{\perp}$  & $\rho_{\parallel}$ & $\rho_{\perp}$ & $\left\langle S^{z}\right\rangle_{\infty}$ \\\hline
square      & 1.4142135      & 1.4142135  & 0.25             & 0.25         & 0.304 \\
triangular  & 1.2990381      & 0.9185586  & 0.2165063        & 0.1082532    & 0.239 \\
maple leaf  & 1.1127356      & 0.6774616  & 0.1584936        & 0.0528312    & 0.154 \\
\end{tabular}
\end{ruledtabular}
\end{table}  
\begin{table}
\caption{The ground state energy per bond $E_{0}/bond$,
the square of sublattice magnetization $m^2$ and the spin gap for
the HAF  on finite 1/7-depleted triangular (maple leaf) lattices
($J=1$). 
For the $N=18$ and
the $N=36$ lattice the results of exact diagonalization are also included. In
the case of the $N=18$ lattice the
variational values were obtained in the whole basis of Ising states, for
larger systems the Monte-Carlo method was applied.
Statistical errors are given in parentheses.}
\label{tab4}
\begin{ruledtabular}
\begin{tabular}{ccccc}
N   &             & $E_0/bond$  & $m^2$      & gap       \\ \hline
18  & exact       & -0.2190     & 0.2832     & 0.5452    \\
    & variational & -0.2083     & 0.2855     & 0.3353    \\ \hline
36  & exact       & -0.2155     & 0.1534     &           \\
    & variational & -0.2027(1)  & 0.179(1)   & 0.146(5)  \\ \hline
72  &             & -0.2001(1)  & 0.128(1)   & 0.071(7)  \\ \hline
162 &             & -0.1991(1)  & 0.099(1)   & 0.020(10) \\ \hline
288 &             & -0.1988(1)  & 0.088(1)   & 0.005(12) \\ \hline
$\infty$&         & -0.1988(2)  & 0.072(1)   &-0.019(25)
\end{tabular}
\end{ruledtabular}
\end{table}               

\end{document}